# The Onset of Fragmentation in Ternary Drop Collisions

Günter Brenn*, Hannes Hinterbichler, Carole Planchette
Institute of Fluid Mechanics and Heat Transfer, Graz University of Technology, Austria
*Corresponding author: guenter.brenn@tugraz.at

**Abstract**
Recently it has been proposed to use colliding drops for producing advanced particles or well defined capsules, or to perform chemical reactions where the merged drops constitute a micro-reactor. For all these promising applications it is essential to determine whether the merged drops remain stable after the collision, forming a single entity, or if they break up. This topic, widely investigated for binary drop collisions of miscible and immiscible liquids, is quite unexplored for ternary drop collisions. The current study aims to close this gap by experimentally investigating collisions of three equal-sized drops of the same liquid, arranged centri-symmetrically. To do this, three drop generators are simultaneously operated to obtain controlled ternary drop collisions. The collision outcomes are observed on photographs and compared to those of binary collisions. Similar to binary collisions, a regime map is built, showing coalescence and bouncing as well as reflexive and stretching separation. Differences are observed in the transitions between these regimes.

**Introduction**
Earliest work on the collisions of pairs of droplets dates back to the late 19[th] century [1]. Interest in the formation of precipitation led to studies on binary water drop collisions in an atmospheric environment in the second half of the 20[th] century [2-4]. Later on, beginning in the 1990s, focus has been set on the investigation of binary collisions of hydrocarbon droplets [5-7]. This was important for understanding the dynamics of sprays with relevance to combustion. In order to determine the influence of viscosity on the droplet collision outcome, many studies with liquids other than water were performed [8-10]. In addition, the ambient gaseous phase was shown to be essential for the collision outcome, especially for bouncing of the drops [7-9]. All the studies mentioned above are limited to the case where the two colliding droplets consist of the same liquid.
New developments in miniaturization and, e.g., advanced drug delivery systems, recently triggered new applications of drop collisions, requiring more than one liquid and/or more than two colliding droplets. Capsules with a liquid core were achieved by colliding pairs of droplets of different liquids [11,12]. In that work, the shell was formed by a solvent exchange process, but it was also proposed to use pairs of immiscible liquids leading to fully liquid capsules [13,14]. In this case, the shell may be hardened after the collision, based, e.g., on in situ polymerization, sol-gel transition or thermal solidification. The production of capsules containing more than one liquid core, or presenting two shells for protecting the contents against environmental influences, may be advantageous for drug delivery. To achieve such architectures, the collision of at least three droplets is needed, as addressed in the present study. Another application of great interest is the use of droplets as micro-reactors. Such an approach has shown high potential for chemical synthesis [15], including the formation of nanoparticles [16] and pharmaceutical applications such as protein analysis [17]. In [15], the droplets were placed on a solid surface and manipulated via electro-wetting, which considerably limits the liquids which may be used. In [16] and [17], as for the majority of droplet micro-reactors [18], the droplets are formed in a microfluidic device requiring the use of an additional immiscible liquid carrier phase and potentially causing insufficient mixing. Replacing sessile and advected droplets by colliding droplets in air could overcome these drawbacks. The flexibility of this process may be enhanced by involving more than two colliding droplets, as it is done in the present work.
In the present study, ternary collisions of droplets consisting of the same liquid are studied experimentally and compared to binary ones with an emphasis on collision outcomes and mechanisms of fragmentation. The collision outcome for constant Ohnesorge number is represented as a function of the non-dimensional impact parameter and the Weber number in collision regime maps for binary and ternary collisions. The drops are made of a glycerol-water solution (50 % glycerol) with an average diameter of 370 µm. Afterwards, the observed differences in the regime boundaries between binary and ternary collisions are discussed and for the case of head-on fragmentation the results of further investigation are presented.

**Material and methods**
The liquids used in the present study were selected such that the properties density, dynamic viscosity and surface tension against air, which are relevant for the collisions, varied over wide ranges. The properties of the liquids – aqueous glycerol solutions of varying concentration and silicon oils – are listed in Table 1. The density $\rho$ was obtained by weighing 5 ml of the liquids, the dynamic viscosity $\mu$ was determined with an Ubbelohde viscometer, and the surface tension $\sigma$ was measured using the pendant drop method with a LAUDA TVT-1 type tensiometer. All the measurements were performed at temperatures of 23±2°C [19].

**Table 1.** Physical properties of the used liquids at 23±2°C. Glycerol concentrations are given in mass percent. [a] Values given by our supplier Carl Roth at 20°C.

| Liquid | Density $\rho$ [kg/m$^3$] | Dynamic viscosity $\mu$ [mPa s] | Surface tension $\sigma$ [mN/m] |
|---|---|---|---|
| Gylcerol 10 % | 1012.46 | 1.23 | 68.94 |
| Glycerol 30 % | 1063.22 | 2.17 | 67.45 |
| Glycerol 40 % | 1096.00 | 3.15 | 66.79 |
| Gylcerol 50 % | 1131.30 | 5.24 | 66.53 |
| Gylcerol 60 % | 1153.88 | 8.81 | 65.27 |
| Gylcerol 65 % | 1166.60 | 13.30 | 64.39 |
| Gylcerol 70 % | 1179.90 | 19.26 | 64.00 |
| Silicon oil m3 | 887.48 | 2.79[a] | 19.50 |
| Silicon oil m5 | 913.40 | 4.57[a] | 19.50 |
| Silicon oil m10 | 931.40 | 9.37[a] | 20.10 |

The experimental setup depicted in Figure 1 allows controlled collisions of three equal-sized droplets to be generated. Streams of monodisperse liquid droplets with diameters of 370±35 µm are obtained from piezoceramic drop generators based on the Plateau-Rayleigh instability [20]. The droplet generators are mounted on translation and rotation stages allowing the adjustment of droplet trajectories with an accuracy of ±2 µm and ±2°, respectively. The liquid is provided by two independent pressurized tanks, the central generator being connected to one tank, the other two to a second one under slightly higher pressure. In this way, the central stream of droplets may be produced with a velocity equal to the vertical component of the velocities of those coming from the sides, so that the collisions may be described in an inertial system moving downstream with that velocity. Typical frequencies of drop formation are in the order of 7 kHz. For illuminating the drops, both an LED lamp synchronized with the drop formation frequency and ultra-short individual flashes (NANOLITE flash lamp) were used. Images of the colliding droplets were recorded with a PCO Sensicam video camera placed on a traverse. The typical resolution of our imaging system is 10 µm/px.

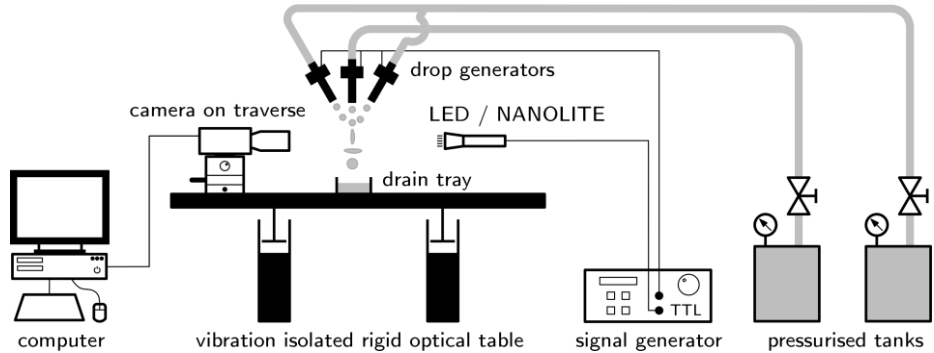

**Figure 1.** Experimental setup for ternary liquid drop collisions. For binary collisions the central drop generator is removed.

The parameters of a collision are obtained from photographs, which is schematically depicted in Figure 2. With knowledge of the projected droplet areas, the drop diameters are calculated by $D = \sqrt{4 \cdot Area/\pi}$. The relative velocities between left and central drop ($\boldsymbol{U_{LC}}$) and right and central drop ($\boldsymbol{U_{RC}}$) are equal in norm ($U_{LC} = U_{RC}$) and opposed in sign ($\boldsymbol{U_{LC}} = -\boldsymbol{U_{RC}}$). Thus, the norm of the relative velocity between left and right drop is

$$U = U_{LC} + U_{RC} = 2U_{LC} = 2U_{RC}. \tag{1}$$

For symmetry reasons the impact parameter between left and central drop ($b_{LC}$) equals the impact between right and central drop ($b_{RC}$). Dividing these two parameters by the droplet diameter leads to the overall non-dimensional impact parameter

$$X = \tfrac{1}{2}(X_{LC} + X_{RC}) = X_{LC} = X_{RC}. \tag{2}$$

The collisions may be fully described by Equation (2) together with two other non-dimensional quantities like Ohnesorge number $Oh = \mu/\sqrt{\rho\sigma D}$ and Weber number $We = \rho U^2 D/\sigma$ (or Reynolds number $Re = \rho UD/\mu$).

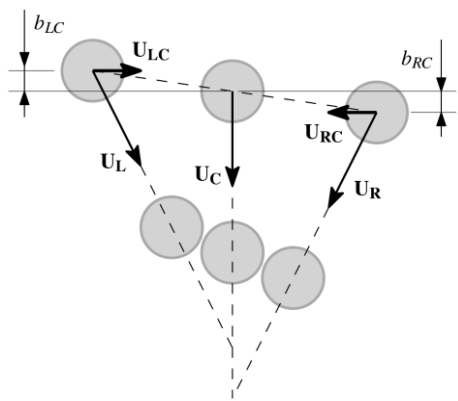

**Figure 2.** Schematic sketch of centri-symmetric ternary drop collision.

**Regimes**

Similar to binary collisions four regimes (excluding droplet shattering at very large Weber numbers), namely *coalescence*, *bouncing*, *reflexive separation* and *stretching separation*, were observed for ternary collisions. These regimes are introduced and discussed below. In the photographs (Figures 3-6) the droplets are seen to move from left to right.

*Coalescence*

In Figure 3 a ternary drop collision at intermediate Weber number resulting in coalescence can be seen. The two outer drops impinge on the central one and deform into a disk shaped complex. After reaching its maximum extension the complex retracts and forms a cylindrical shape droplet. Due to surface tension forces the cylinder relaxes into a spherical droplet (farther downstream, not shown), dissipating its excess surface energy due to viscosity during subsequent oscillations. At collisions with small Weber number the merged complex is less deformed and, as a consequence, regains its spherical shape faster. Coalescence can also occur at collisions with nonzero impact parameter (not shown). In this case a rotation is induced into the merged complex.

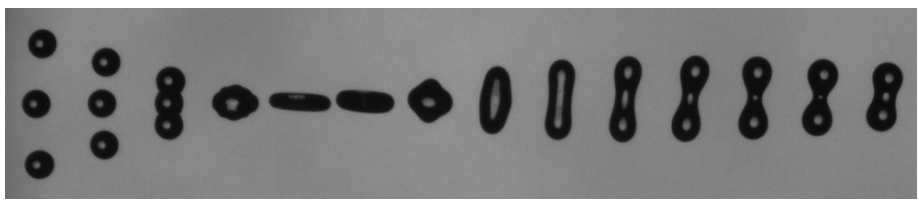

**Figure 3.** Coalescence after a ternary head-on collision: $D$=357 µm, $We$=44.3, $X$=0.01.

*Bouncing*

The phenomenon of bouncing occurs at rather large non-dimensional impact parameters. It is exemplarily shown in Figure 4 for $We$=47.2 and $X$=0.63. The intervening air layer cannot be expelled and the droplets bounce apart since no liquid bridge can be established. As we will show later on, bouncing can be observed over a wide range of Weber numbers.

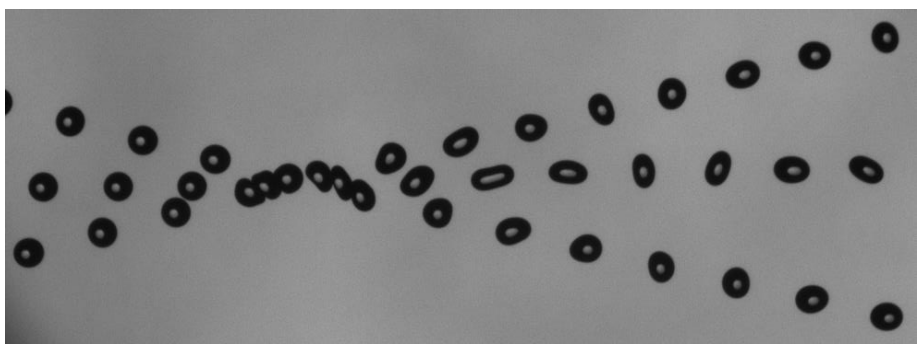

**Figure 4.** Bouncing drops at a ternary collision: $D$=369 µm, $We$=47.2, $X$=0.63.

*Reflexive separation*

Reflexive separation of three colliding droplets can be seen in Figure 5. Very similar to coalescence in Figure 3, the droplets form a disk shaped complex after the impact, which relaxes into a cylindrical rod. However, in Figure 5, the relaxation drives the extremities of this cylinder farther outwards. Eventually the cylinder breaks into two main droplets, similar to what is observed for Rayleigh instability of infinitely long liquid cylinders. For large $We$, a different fragmentation process is observed (not shown). In that case, after the relaxation of the disk, a very long cylindrical rod, with its length exceeding its diameter many times, is formed. Due to the *end-pinching* mechanism [21-22], droplets pinch off both extremities resulting in at least three droplets of approximately the same size. Stretching separation can be observed for head-on collisions and small impact parameters ($X < 0.2$). At a larger non-dimensional impact parameter the collision would result in stretching separation, which is discussed next.

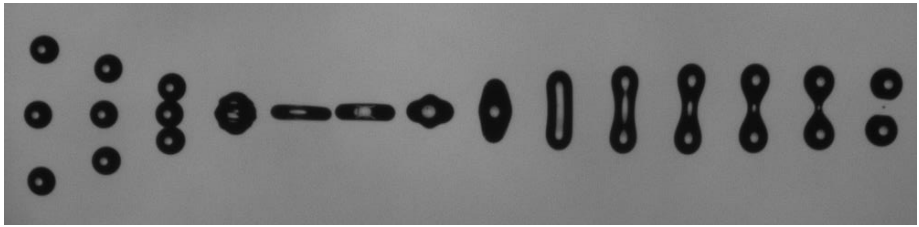

**Figure 5.** Reflexive separation after a ternary head-on collision: $D$=359 µm, $We$=48.3, $X$=0.00.

*Stretching separation*

An example of stretching separation is depicted in Figure 6 at $We$=90.6 and $X$=0.44. Although the three droplets merge, due to inertia, the outer droplets remain on their initial trajectories. Thus, a connecting ligament is formed which is stretched until the outer drops are pinched off. The remaining ligament subsequently breaks up into several satellite droplets due to *end-pinching* mechanism (not shown in Figure 6, occurs farther downstream). An increase of $X$ would lead to bouncing, a decrease to stretching separation. Reflexive separation cannot be observed for head-on collisions.

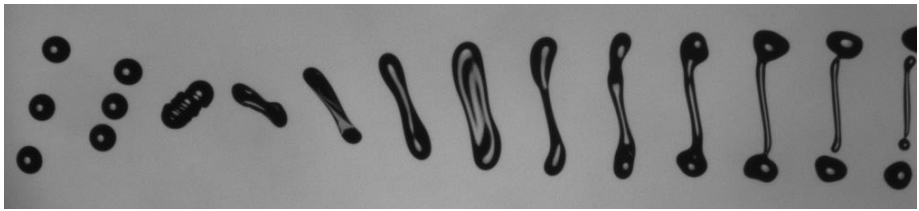

**Figure 6.** Reflexive separation after a ternary collision: $D$=386 µm, $We$=90.6, $X$=0.44.

**Map of regimes**

The collision regime map for centri-symmetric ternary drop collisions of equal sized droplets is obtained by plotting the collision outcomes at constant $Oh$ and varying $We$ and $X$ into a single nomogram. The map of regimes at an average Ohnesorge number $Oh$=0.031 (using glycerol 50% as the liquid) can be seen in Figure 7 on the left hand side. Since perfect symmetric collisions cannot be achieved in practice, each data point is plotted with error bars in terms of $We$ and $X$. The uncertainty in $We$ is a result of the variability of the relative velocity between $2U_{LC}$ and $2U_{RC}$ (see Equation (1)), whereas the non-dimensional impact parameter is plotted between its two extreme values $X_{LC}$ and $X_{RC}$ (see Equation (2)).

The boundaries between the four regimes, which were described above, are empirically drawn as grey lines. At small Weber numbers (between 10 and 45) only the regimes of coalescence and bouncing can be observed, the former at small and medium impact parameters and the latter above a critical impact parameter. At a Weber number of approximately 45 the regime of reflexive separation appears for head-on collisions and stretching separation occurs at medium impact parameters ($X \approx 0.5$). With increasing $We$ the transition from stretching separation to bouncing is shifted to larger $X$, whereas the opposite is true for the transition from stretching separation to coalescence. At $We \approx 90$ the regime of coalescence cannot be observed any more. For Weber numbers larger than 90 investigated in the present study the transition between reflexive and stretching separation propagates at an almost constant non-dimensional impact parameter of $X \approx 0.22$.

As a next step, in order to gain further insight into ternary drop collisions, we compare our results to binary drop collisions. Therefore, we obtained a collision regime map at the same Ohnesorge number (same droplet size and same liquid) as in the ternary case, which can be seen in Figure 7 on the right hand side. For binary collisions no error bars are necessary since only one impact parameter and relative velocity occur in this case. Again, the transitions are empirically drawn as grey lines. Differences and similarities between binary and ternary drop collisions are discussed in the following section.

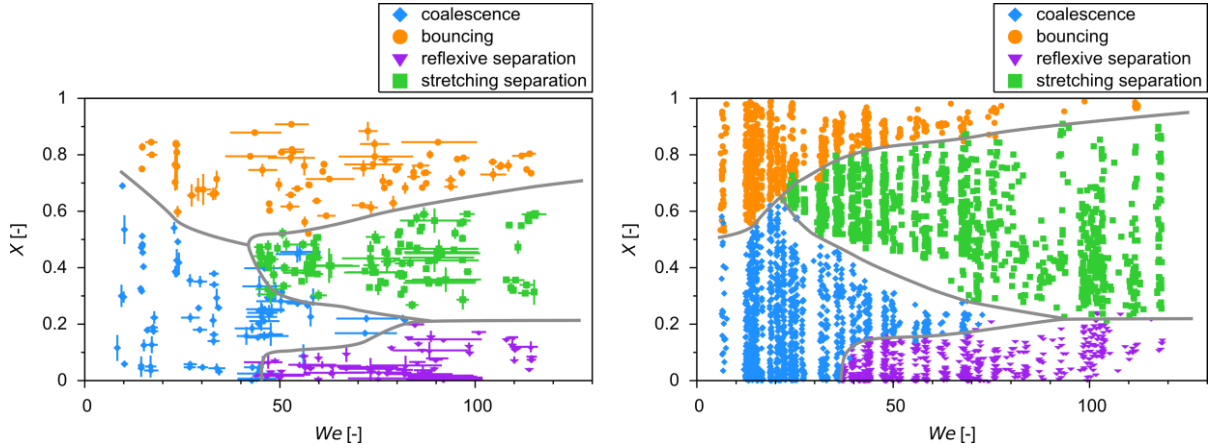

**Figure 7.** Left: Map of regimes for ternary drop collisions (centri-symmetric) of equal sized droplets. The used liquid is glycerol 50%. The average drop diameter is 370 µm resulting in an Ohnesorge number of $Oh$=0.031. The Weber number ranges from 8 to 115. The error bars for $X$ are drawn between $X_{LC}$ and $X_{RC}$ (see Equation (2)) and the error bars for $We$ are drawn with respect to $2U_{LC}$ and $2U_{RC}$ (see Equation (1)). Right: Map of regimes for binary drop collisions using glycerol 50% as the liquid. The average drop diameter is 350 µm resulting in $Oh$=0.032. The Weber number ranges from 6 to 120.

**Transitions between regimes**

In the present section the transitions between the observed regimes are compared for binary and ternary drop collisions, with an emphasis on differences between these two types of collisions. Thus, the transitions between coalescence and stretching separation, as well as between reflexive and stretching separation, are excluded from the discussion since they agree to a great extent [19]. In Figure 8 the regime boundaries for both binary and ternary drop collisions are drawn according to Figure 7.

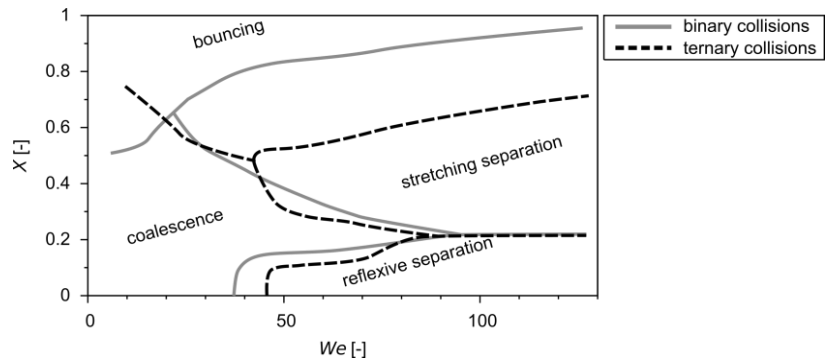

**Figure 8.** Regime boundaries for binary and ternary drop collisions according to Figure 7.

*Bouncing – coalescence and bouncing – stretching separation*

In Figure 8 significant differences can be observed in the transition between stretching separation and bouncing, and between coalescence and bouncing. The former (stretching separation – bouncing) is shifted to lower non-dimensional impact parameters for ternary collisions with an almost constant offset of approximately 0.2, whereas the latter (coalescence – bouncing) rises with increasing Weber number for binary collisions and declines for ternary collisions.

These differences may be explained by considering ternary bouncing rather as two binary collisions at the same time than a single ternary collision [19]. Since the outer drops never directly interact (see Figure 4) the mechanisms of binary and ternary bouncing can be assumed equivalent. As a matter of fact, this assumption appears to be practical, which can be seen below. In Figure 9, the Weber numbers of the data points representing ternary bouncing are calculated with $U/2$, which is in the same order of magnitude as $U_{LC}$ and $U_{RC}$ (see Equation (1)), corresponding to binary collisions. Thus, the initial ternary transition (black dashed line) is replaced by the one marked with a red solid line in Figure 9. It can be seen that the transitions for binary and ternary collisions now agree very well. Only at very small Weber numbers close to zero some deviations can be observed. They can be explained by the experimental uncertainty of $We$ and $X$ in this range (see Figure 7).

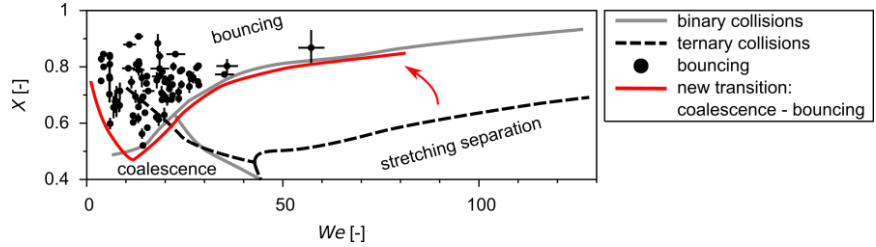

**Figure 9.** Regime boundaries for binary and ternary liquid drop collisions, according to Figure 7. The non-dimensional impact parameter is plotted with and error bar between $X_{LC}$ and $X_{RC}$ (see Equation (2)). The Weber number is calculated with $U/2$ with an uncertainty corresponding to $U_{LC}$ and $U_{RC}$.

*Coalescence – reflexive separation*

Looking at Figure 8, it appears that for head-on collisions the transition between coalescence and reflexive separation can be found at smaller Weber numbers for binary than for ternary collisions.

To further compare these two cases, and to stress similarities and differences between them, we focus (i) on the first phase of the collision and (ii) on the last instants preceding fragmentation, considering a wide range of liquid properties, drop sizes and relative velocities.

The first phase (i) extends from the drop contact (taken as time origin) to the instant t_Dmax when the transient disk-shaped entity has reached its maximal diameter (typically corresponding to the 5th triplet from the left in Figure 3). Making use of the analogy between a drop and a liquid spring, this phase can be seen as the spring's compression [14]. We extracted from the photographs, $t_{Dmax}$ for both the binary and ternary collisions and compared it to $t_{oscill}$ defined by $\sqrt{\rho D^3/\sigma}$ and corresponding to a fraction of a drop oscillation period. Our results are presented in Figure 10 together with the results of Willis & Orme [9] for the binary collisions. In Figure 10a, $t_{Dmax}/t_{oscill}$ is plotted as a function of $We$. For large enough We (typically $We>50$), we observe that $t_{Dmax}/t_{oscill}$ becomes constant, ruling out a possible scaling in $D/U$. In Figure 10b, $t_{Dmax}$ is plotted as a function of $t_{oscill}$ for points corresponding to $We>50$. The proportionality between $t_{Dmax}$ and $t_{oscill}$ is confirmed, validating the analogy with compressing springs. The ratio of the slopes for binary and ternary cases (0.32 and 0.49, respectively) is 0.65~2/3, as expected considering the association of 2 and 3 springs in series, for binary and ternary collisions, respectively. Phenomenologically, the first phase of the collisions appears very similar and can be seen as the compression of liquid springs for both binary and ternary cases.

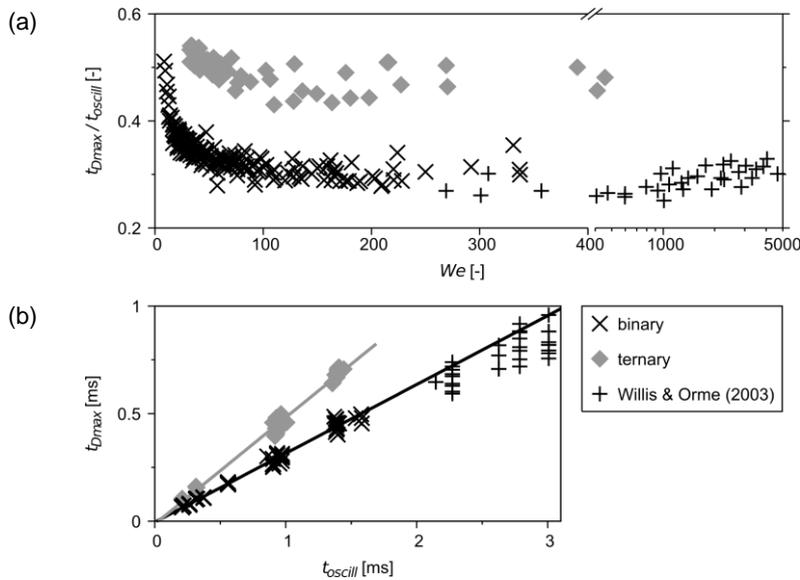

**Figure 10.** (a) $t_{Dmax}/t_{oscill}$ as a function of We for binary (x our data, + Willis & Orme data [9]) and ternary collisions. (b) $t_{Dmax}$ as a function of $t_{oscill}$ for points corresponding to $We>50$. The slopes are 0.32 and 0.49 for binary and ternary collisions, respectively.

Further comparison of the binary and ternary drop collisions can be made by looking at the last instants preceding the fragmentation (ii). At this stage the merged liquid entity is shaped as a cylinder, and its aspect ratio $\zeta$ can be measured from photographs (maximum aspect ratio typically at 5th triplet on the right of figure 5). Here $\zeta$ is defined as $L/d$ where $d$ is the cylinder diameter and $L$ its length. Approaching the threshold velocity, it is possible to estimate $\zeta_{crit}$, the critical value of the aspect ratio leading to fragmentation. In practise, the upper and lower bounds of $\zeta_{crit}$ are taken as the last value obtained for coalescing drops and the first one found for fragmenting drops, as done in [14] for immiscible liquids. The results are plotted in Figure 11 as a function of $Oh$ for both binary and ternary collisions.

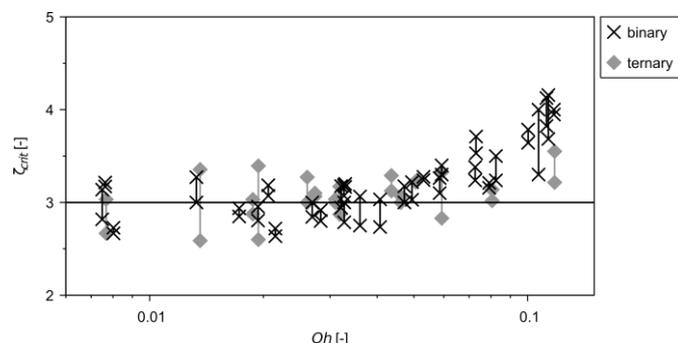

**Figure 11.** $\zeta_{crit}$ as a function of $Oh$ for binary and ternary collisions

We observe that below a certain value of $Oh$ (typically up to 0.1) $\zeta_{crit}$ is constant, close to a value of 3, for both binary and ternary collisions. Note that the value of 3 is close to the theoretical value of $\pi$ found for the classical Rayleigh criterion applied to an infinite liquid cylinder at rest.

These analyses (i) and (ii) tend to indicate that binary and ternary collisions are very similar. In both cases it is relevant to consider drops as liquid springs that compress, merge and relax. The fragmentation criterion can be approached by the Rayleigh criterion, and differences between binary and ternary collisions are therefore expected to be mainly limited to velocity fields and gradients within the merged entity.

To assess impacts of potential differences on fragmentation thresholds, we have plotted in Figure 12 $We^*$ as a function of $Oh$. Here, $We^*$ represents the threshold ratio of initial kinetic and surface energies, defined by $We_c/48$ and $We_c/72$ for binary and ternary cases, respectively. $We_c$ is the Weber number for which the transition between coalescence and reflexive separation is found. For low $Oh$ (typically $Oh<0.04$), $We^*$ is found to be very similar for both configurations. We interpret this finding as the fact that, for low $Oh$, drop collisions happen in a capillary-inertial regime. Viscous losses are limited to boundary layers, and due to the similarities of binary and ternary collisions, the driving parameter for fragmentation is $We^*$, the ratio between initial kinetic and surface energies. For higher $Oh$ (typically $Oh>0.06$), $We^*$ obtained for binary collisions becomes significantly larger than We* of ternary ones. For a given $Oh$, the ratio between $We^*$ approaches 3/2, indicating that threshold velocities become similar. We interpret this as the consequence of a change from a capillary-inertial regime to a viscous-inertial regime [14]. Viscous losses are not limited to boundary layers, but develop in the whole liquid volume. The driving parameter for fragmentation is thus the Reynolds number.

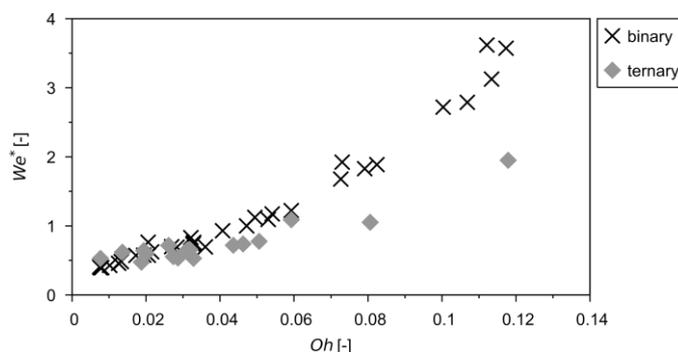

**Figure 12.** $We^*$ as a function of $Oh$ for binary and ternary collisions.

**Conclusions**
Ternary drop collisions were investigated for their relevance in micro-reaction and particle technologies. A survey of the mechanisms governing the outcomes from the collisions shows similarities to the binary case, seen in similar boundaries between regimes of collision outcomes in a map of the non-dimensional impact parameter and the collision Weber number. The basic mechanisms of coalescence, bouncing, reflexive separation and stretching separation are observed in both cases. A closer look into the regime boundaries, however, reveals differences. It is found that the transition from stretching separation to bouncing at high non-dimensional impact parameter may be reduced to the behaviour of two binary collisions going on in the ternary case. Analysing the colliding drops as elastic springs compressed in the first phase after the impact, the time $t_{Dmax}$ elapsed until maximum deformation is reached may be compared to a time scale $t_{oscill}$ characteristic for an elastic spring, with the surface tension as the spring constant. With increasing impact Weber number, the ratio of these two times converges to a constant value, which is different in the binary and the ternary cases. The ratio of the slopes of $t_{Dmax}$ against $t_{oscill}$ for the binary and ternary cases is close to 2/3, which is to be expected for the colliding drop systems interpreted as associations of 2 and 3 springs. Differences between binary and ternary collisions are also seen in the dependency of the critical collision Weber number $We^*$ found for different regimes of the drop Ohnesorge number. In the capillary-inertial regime of low Ohnesorge number, $Oh$<0.04, critical Weber numbers for binary and ternary collisions come out very similar. For higher $Oh$>0.06, in the viscous-inertial regime, the critical Weber number is significantly larger for binary than for ternary collisions. At the high $Oh$, viscous losses are not limited to boundary layers, but develop in the whole liquid volume. The number characterising the state critical for fragmentation is therefore the Reynolds number.

**Acknowledgements**
H.H. acknowledges support from the Reseach Center Pharmaceutical Engineering at Graz University of Technology. Participation in the EU COST action MP1106 is gratefully acknowledged.

**Nomenclature**

| | | |
|---|---|---|
| $b$ | impact parameter [m] | |
| $d$ | cylinder diameter [m] | |
| $D$ | drop diameter [m] | |
| $L$ | cylinder length [m] | |
| $Oh$ | Ohnesorge number [-] | |
| $Re$ | Reynolds number [-] | |
| $t$ | time [s] | |
| $\boldsymbol{U}$ | relative velocity vector [m/s] | |
| $U$ | norm of relative velocity vector [m/s] | |
| $We$ | Weber number [-] | |
| $X$ | non-dimensional impact parameter [-] | |
| $\zeta$ | aspect ratio of the cylinder [-] | |
| $\mu$ | dynamic viscosity [Pa s] | |
| $\rho$ | density [kg/m$^3$] | |
| $\sigma$ | surface tension [N/m] | |

**Subscripts**

| | |
|---|---|
| $c$ | transition between coalescence and reflexive separation |
| $crit$ | critical value leading to fragmentation |
| $Dmax$ | maximum deformation of the disk |
| $LC$ | between left and central drop |
| $oscill$ | oscillation period |
| $RC$ | between right and central drop |